\documentclass[journal]{IEEEtranTIE}

\usepackage[utf8]{inputenc}     
\usepackage[T1]{fontenc}
\usepackage{graphicx}
\usepackage{amsmath}
\usepackage{amssymb}
\usepackage{epstopdf}
\usepackage{flushend}
\usepackage[hidelinks]{hyperref}
\usepackage{soul}           
\usepackage{subfigure}
\usepackage{float}
\usepackage{multicol}
\usepackage[most]{tcolorbox}
\usepackage{cite}
\usepackage{multirow}
\usepackage[figuresright]{rotating}
\usepackage{placeins}
\usepackage{cancel}
\usepackage{url}

\usepackage{amsmath}
\usepackage{algorithm}
\usepackage[noend]{algpseudocode}

\usepackage{stfloats}
\usepackage[dvipsnames]{xcolor}

\usepackage{booktabs}

\usepackage{tikz}
\usetikzlibrary{positioning, arrows.meta, calc, decorations.pathreplacing, calligraphy, fit, backgrounds}

\newcommand{\sinalpha}{\sin{\alpha}}
\newcommand{\sinbeta}{\sin{\beta}}
\newcommand{\cosalpha}{\cos{\alpha}}
\newcommand{\cosbeta}{\cos{\beta}}

\newcommand{\fricfunX}{f_x(x_t,\dot{x}_t)}
\newcommand{\fricfunY}{f_y(y_t,\dot{y}_t)}

\newcommand{\fricfunXbar}{\bar{f}_x(x_t,\dot{x}_t)}
\newcommand{\fricfunYbar}{\bar{f}_y(y_t,\dot{y}_t)}
\newcommand{\fricfunlbar}{\bar{f}_l(\dot{L})}

\newcommand{\supermenos}{\scriptscriptstyle-}
\newcommand{\supermas}{\scriptscriptstyle+}
\newcommand{\supermasmenos}{\scriptscriptstyle\pm}

\begin{document}

\title{A hybrid dynamic model and parameter estimation method for accurately simulating overhead cranes with friction.}

\author{
	\vskip 1em
    Jorge Vicente-Martinez and~Edgar Ramirez-Laboreo

    \thanks{
    
		\phantom{{\color{red}Manuscript received dd Month 2xxx; revised dd Month 2xxx; accepted dd Month 2xxx. Date of publication dd Month 2xxx; date of current version dd Month 2xxx.}}

        \vspace{8mm}
        
        This work is part of the project \mbox{PID2024-159279OB-I00}, funded by MICIU/AEI/10.13039/501100011033 and by ERDF/EU. It was founded also in part by the MICIU through the grant FPU24/01878, in part by the Government of Arag\'on - EU, via grant T45{\_}23R and in part by Fundaci\'on Ibercaja and the University of Zaragoza, via grant \mbox{JIUZ2023-IA-07}. \textit{(Corresponding author: J. Vicente-Martinez.)}

        \vspace{2mm}
        
        The authors are with the Departamento de Informatica e Ingenieria de Sistemas (DIIS) and the Instituto de Investigacion en Ingenieria de Aragon (I3A), Universidad de Zaragoza, 50018 Zaragoza, Spain (e-mail: j.vicente@unizar.es; ramirlab@unizar.es).
            
        \phantom{{\color{red}Color versions of one or more of the figures in this paper are available online at XXXXXXXXXXXXXXXX.}}

        This article has supplementary downloadable material available at \url{https://github.com/jrgvicente/Crane3DSim}, provided by the authors.
		
        \phantom{{\color{red}Digital Object Identifier XX.XXXX/XXX.20XX.xxxxxxx}}
        
    }
        
}


{}


\maketitle

\begin{abstract}

This paper presents a new approach to accurately simulating 3D overhead cranes with friction. Although nonlinear friction dynamics has a significant impact on these systems, accurately modeling this phenomenon in simulations is a significant challenge. Traditional methods often rely on imprecise approximations of friction or require excessive computational times for reliable results. To address this, we present a hybrid dynamical model that features a trade-off between high-fidelity friction modeling and computational efficiency.
Furthermore, we present a step-by-step algorithm for the comprehensive estimation of all unknown system parameters, including friction. This methodology is based on Bayesian Linear Regression and Least Squares (LS) estimations. Finally, experimental validation with a laboratory crane confirms the effectiveness of the proposed modeling and estimation approach.

\end{abstract}


% %%%%%%%%%%%%%%%%%%%%%%%%%%%%%%%%%%%%%%%%%%%%%%%%%%%%%%%%%%%%%%%%%%%%%%%%%%%%%%
% \section{Introduction}
% %%%%%%%%%%%%%%%%%%%%%%%%%%%%%%%%%%%%%%%%%%%%%%%%%%%%%%%%%%%%%%%%%%%%%%%%%%%%%%

\IEEEPARstart{O}{verhead} cranes are essential equipment in modern industrial environments for lifting and transporting heavy loads across diverse sectors such as construction, logistics, and shipbuilding. The widespread use of these devices means that their automation and control remains a prominent field of research \cite{li_rrt_2026,wei_fixedtime_2025}. Currently, crane operation relies heavily on manual control, as achieving precise control of the position of the payload to ensure safe and efficient operation is not trivial due to the nonlinear dynamics relating the control inputs to the motion of the load.

To address this challenge, researchers have developed a wide range of controllers that primarily focus on suppressing load oscillations while controlling the position of the payload during crane operation. A comprehensive overview of modeling and control of overhead cranes over recent years is available in \cite{mojallizadeh_modeling_2023}. Furthermore, recent works continue to develop anti-swing controllers using different approaches, such as \cite{zhou_nonobserver-based_2025}, where the authors propose a fixed-time feedback controller that achieves fast load swing reduction without using observers.

However, to implement feedback control strategies, many works design controllers using simplified models that either treat friction as a disturbance \cite{wu_disturbance-observer-based_2020, ma_neural_2022} or neglect it to focus on the theoretical development of position and no-oscillation feedback controllers \cite{moustafa_nonlinear_1988, lee_modeling_1998}. Furthermore, these models are frequently used to validate controller performance in simulation, neglecting real-world friction dynamics. Yet, the critical influence of friction is well-documented in other mechanical systems like robotic manipulators \cite{huang_adaptive_2025} and pendular systems \cite{soto_modelado_2020}. In the overhead crane context, recent studies also highlight the necessity of addressing friction, proposing approaches such as neural network-based compensation \cite{zhu_two-phase_2025}. To accurately capture these effects, one of the most widely used friction models is a superposition of dry (Coulomb) and viscous friction. This structure is frequently employed in the modeling of robotic and mechatronic systems to capture both the static and dynamic dependencies of friction \cite{bona_friction_2005}.

 The physical nature of dry friction is usually modeled as a constant that depends on the direction of motion. This introduces a discontinuity at zero velocity. In simulations, this situation causes an undesired chattering effect, resulting in a friction force that alternates between positive and negative values \cite{cisneros_reliable_2020}. This forces simulation solvers to drastically reduce their integration step-size at zero-crossings, thereby increasing the computational cost. 
 To overcome this issue, a widespread practice is to use approximations of this dry friction function with continuous and differentiable functions, with the hyperbolic tangent being one of the most popular choices \cite{wenzl_comparison_2018}.
 This approach, in its various forms, has been used in several studies on overhead cranes that do consider friction in their simulations \cite{jaafar_control_2021, bello_modelling_2024, ma_adaptive_2008, guo_anti-swing_2023}.
 The main drawback of this approximation is that, in order to faithfully emulate the discontinuity of the Coulomb model, this function must be approximated as much as possible to the $\mathrm{sign}$ function. This slows down the simulations and can compromise the feasibility on applications that need fast simulations.

 \IEEEpubidadjcol

Various alternative models have been proposed to avoid these discontinuity issues, such as the continuously differentiable function in \cite{makkar_lyapunov-based_2007}. Other works, such as \cite{kolar_time-optimal_2017}, propose a nonlinear state observer for friction compensation but still rely on smooth approximations of the $\mathrm{sign}$ function.

 In addition to the correct representation of the zero-crossing behavior, an accurate friction model for overhead cranes must consider other particularities. As highlighted in \cite{khatamianfar_new_2014} and \cite{lobe_flatness-based_2018}, the dry and viscous friction coefficients can exhibit a significant dependence on the direction of movement and the position of the trolley along the rail. Some works in other areas have shown that estimating all friction parameters simultaneously is challenging due to the high correlation between them \cite{wenzl_comparison_2018}. In overhead cranes, \cite{khatamianfar_new_2014} proposes a physical parameters estimation method including dry friction, but the position dependence is not taken into account. In \cite{lobe_flatness-based_2018}, the authors use a Least Squares (LS) identification procedure for position-- and direction--dependent dry friction; however, the estimation process is not explained.

 While simulations with a less accurate friction model may be sufficient to demonstrate the correct operation of some controllers, other applications, such as the design of flatness-based feedforward controllers or reinforcement learning-based approaches, can benefit significantly from including additional details in the model used for both modeling and simulation. 
 In \cite{kolar_time-optimal_2017}, the authors present experimental results comparing the performance of a laboratory crane under feedforward control using a model with and without friction. They conclude that feedforward controllers can be improved significantly including dry friction compensation. Thus, to verify this behavior in simulation, the model must include this phenomenon.

Articles developing overhead crane controllers often use different models, which makes direct comparison of controllers challenging. We aim to provide the scientific community with a lightweight and scalable model that will allow researchers to rigorously test and compare their control strategies. The code for this simulator is available at GitHub, on the provided supplementary material link.

In summary, the main contributions of this work are:
\begin{itemize}
    \item The development of a hybrid simulation model, combining both continuous and discrete dynamics, that allows performing fast simulations while maintaining the accuracy of the ideal discontinuous Coulomb model. We include comparisons between the computational time performance of the approximated model and our model in some illustrative cases.
    \item The analysis of friction dependence on position and direction with real data, and the inclusion of this dependence into the simulation model.
    \item The development of a step-by-step algorithm for parameter estimation of the hybrid model that includes position and direction dependence of friction.
    \item The validation of this estimation method and the hybrid model with real data from a laboratory overhead crane.
    
\end{itemize}

%%%%%%%%%%%%%%%%%%%%%%%%%%%%%%%%%%%%%%%%%%%%%%%%%%%%%%%%%%%%%%%%%%%%%%%%%%%%%%
\section{System modeling}
\label{sec:modelo}
%%%%%%%%%%%%%%%%%%%%%%%%%%%%%%%%%%%%%%%%%%%%%%%%%%%%%%%%%%%%%%%%%%%%%%%%%%%%%%

%----------------------------------------------------------------------------%
\subsection{System description}
\label{sec:descripcion}
%----------------------------------------------------------------------------%

\begin{figure*}
\centering
    \includegraphics[width=17cm]{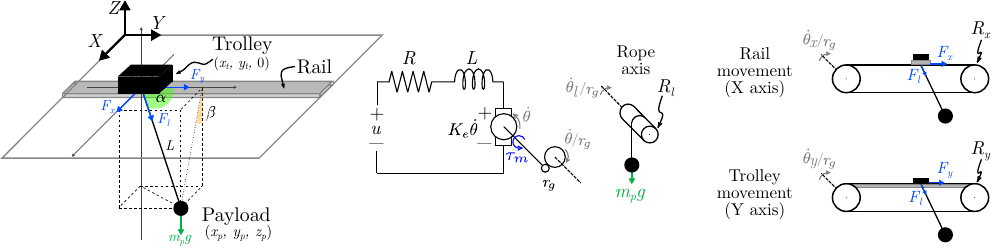}
    \caption{Schematic diagram of the overhead crane, one DC motor and the forces on the 3 axes.}
    \label{figdiagrama}
\end{figure*}

The system under consideration is a three-dimensional overhead crane, and while the developed model is general, in this paper it is particularized for a setup consisting of three DC motors and a trolley-payload system, as shown schematically in Fig.~\ref{figdiagrama}.

The mechanical assembly consists of these components:
\begin{itemize}
    \item A rail of mass $m_r$ that moves along the X-axis over the lateral beams of the bridge, driven by force $F_x$.
    \item A trolley of mass $m_t$ that moves along the Y-axis on the rail, driven by a force $F_y$.
    \item A payload of mass $m_p$ hangs from the trolley by means of a rope that can change length $L$ along its axis. The hoisting mechanism provides a lifting force $F_l$.
\end{itemize}

The actuation forces $F_x$, $F_y$, and $F_l$ are produced by three independent DC motors, which are driven by input voltages $u_x$, $u_y$, and $u_l$, respectively. Each motor is coupled to its corresponding axis (rail, trolley, and hoist) via a gearbox and a pulley.
The relation between the linear forces and the effective torques ${\tau_{c}}_{x}, {\tau_{c}}_{y}, {\tau_{c}}_{l}$ are
\begin{equation}
\label{eq: torque_lineal relacion}
      {\tau_{c}}_{x} = \frac{R_x}{r_g}\,F_x \;\;,\;\; {\tau_{c}}_{y} = \frac{R_y}{r_g}\,F_y   \;\;,\;\;     {\tau_{c}}_{l} = \frac{R_l}{r_g}\,F_l , 
\end{equation}
where $r_g$ is the gearbox transmission ratio and $R_x$, $R_y$, and $R_l$ are the radii of the drive pulleys for the X, Y, and rope axes.

Similarly, the kinematic relationships between the linear displacements of the trolley and the rope length, $x_t$, $y_t$, $L$, and the angular positions of the motor axes $\theta_x$, $\theta_y$, $\theta_l$ are
\begin{equation}
\label{eq: velocidad_lineal relacion}
    {\theta}_x = \frac{r_g}{R_x}\,{x_t}  \;\;,\;\; {\theta}_y = \frac{r_g}{R_y}\,{y_t} \;\;,\;\; {\theta}_l = \frac{r_g}{R_l}\,{L}.
\end{equation}

%----------------------------------------------------------------------------%
\subsection{DC Motors dynamics}
%----------------------------------------------------------------------------%
The following equations describe the X-axis motor dynamics. The equations for the Y and rope axes are equivalently defined.

By assuming the electrical dynamics are negligible compared to the mechanical dynamics, the motor's inductance is omitted. The resulting generic electrical equation for the X-axis DC motor is given as follows:
\begin{equation}
\label{eq: motor1}
    u_x=R i_x + K_e\dot{\theta}_x \,,
\end{equation}
where $u_x$ is the input voltage, $i_x$ is the electric current and $R$ is the internal motor resistance.
The electromotive force is related to the angular velocity via the constant $K_e$.
The torque balance equation for the motor is
\begin{equation}
    {\tau_m}_x = J_x \ddot{\theta}_x +{\tau_c}_x+{{\tau}_f}(\dot{\theta}_x)\,,%% Le pongo D a la viscosa y C a la seca (de coulomb)
\end{equation} 
where ${\tau_m}_x$ is the motor-generated torque and $J_x$ is the motor inertia.
Consequently, the motor torque is distributed between overcoming the rotor's inertia, compensating for friction, ${\tau}_f(\dot{\theta}_x)$, and driving the external load.
The torque generated by the motor is proportional to the current via the motor torque constant, $K_p$:
\begin{equation}
\label{eq: motor3}
    {\tau_m}_x = K_p i_x\,.
\end{equation}

%----------------------------------------------------------------------------%
\subsection{Overhead crane motion dynamics}
%----------------------------------------------------------------------------%

The position of the payload ($x_p$, $y_p$, $z_p$) can be determined from the position of the trolley, $x_t, y_t$, the length of the rope, $L$, and the swing angles $\alpha$ and $\beta$ (see Fig. \ref{figdiagrama}):
\begin{align}
    x_p &= x_t + L\sinbeta\,\sinalpha\,, \\
    y_p &= y_t + L\cosalpha\,, \\
    z_p &= -L\sinalpha\cosbeta\,.
\end{align}

Friction forces between trolley-rail and rail-beams are included in the model with the functions $\fricfunX$ and $\fricfunY$.
Friction between the payload and the air has not been considered on the rope axis. 
However, we have included viscous friction in the dynamics of angles $\alpha$ and $\beta$ 
based on measurements obtained from the real system where the friction is produced by the crane swing system assembly itself.

Using the Lagrangian method \cite{lee_modeling_1998}, the equations of motion for each degree of freedom are derived:
\begin{equation}
\label{eq:x_carro_solo_grua}
\begin{split}
\ddot{x}_t \,(m_t + m_r) &= F_x - \fricfunX - F_l \sinalpha \,\sinbeta \\ 
           &\quad + \tfrac{D_\alpha}{L}\dot{\alpha}\cosalpha\,\sinbeta +\tfrac{D_\beta}{L\sinalpha}\dot{\beta}\,\cosbeta \,,
\end{split}
\end{equation}
\begin{equation}
\label{eq:y_carro_solo_grua}
\ddot{y}_t \,m_t = F_y - \fricfunY - F_l \cosalpha - \tfrac{D_\alpha}{L}\dot{\alpha}\,\sinalpha \,,
\end{equation}
\begin{equation}
\label{eq:L_solo_grua}
\begin{split}
\ddot{L}\, m_p &= F_l + L \,\dot{\alpha}^2 \,m_p - \ddot{y}_t m_p\cosalpha \\
         &\quad+ L \dot{\beta}^2 \,m_p\sin^2{\!\alpha} - \ddot{x}_t m_p\sinalpha \,\sinbeta  \\
         &\quad+ m_p g \cosbeta \,\sinalpha  \,,
\end{split}
\end{equation}
\begin{equation}
\label{eq:dyn_alpha}
\begin{split}
\ddot{\alpha}\, L &= \Big( \ddot{y}_t \sinalpha  - \ddot{x}_t \cosalpha \,\sinbeta  + g\cosalpha\, \cosbeta  \\
              &\quad + L\dot{\beta}^2\cosalpha \,\sinalpha - 2\dot{L}\dot{\alpha} \Big) - \frac{D_{\alpha} \dot{\alpha}}{m_p L} \,,
\end{split}
\end{equation}
\begin{equation}
\label{eq:dyn_beta}
\begin{split}
\ddot{\beta}\, L\sinalpha &= -g\sinbeta - \cosbeta \ddot{x}_t - 2\sinalpha \dot{L}\dot{\beta} \\
             &\quad- 2\cosalpha L\dot{\alpha}\dot{\beta}  - \frac{D_{\beta} \dot{\beta}}{m_p L \sinalpha} \,,
\end{split}
\end{equation}
where $g$ is the gravity constant and $D_\alpha$ and $D_\beta$ are, respectively, the viscous friction coefficients in the $\alpha$ and $\beta$ angles.
Although the system is implicit, the explicit equations can be obtained by solving for $\ddot{x}_t$ and $\ddot{y}_t$ in \eqref{eq:x_carro_solo_grua}--\eqref{eq:y_carro_solo_grua} and substituting into \eqref{eq:L_solo_grua}--\eqref{eq:dyn_beta}.

%----------------------------------------------------------------------------%
\subsection{Complete dynamics}
%----------------------------------------------------------------------------%

By particularizing the DC motor equations \eqref{eq: motor1}--\eqref{eq: motor3} for each axis and integrating them with the mechanical dynamics of the crane \eqref{eq:x_carro_solo_grua}--\eqref{eq:dyn_beta}, the complete continuous system 
\begin{equation}
    \mathbf{\dot{x}} = g( \mathbf{x}, \mathbf{u})
\end{equation}
is obtained, where $ \mathbf{\dot{x}}$ is the state vector and $\mathbf{u}$ the input vector
\begin{align}
    \mathbf{x} &= [\;x_t \;\; \dot{x}_t \;\; y_t \;\; \dot{y}_t \;\; L \;\; \dot{L} \;\; \alpha \;\; \dot{\alpha} \;\; \beta \;\; \dot{\beta}\;]^\intercal \,,
    \label{eq: xpunto_g_x_u} \\
    \mathbf{u} &= [\;u_x \;\; u_y \;\; u_l]\,.
\end{align}

The complete dynamics of \eqref{eq: xpunto_g_x_u} is implicitly described by \eqref{eq:dyn_alpha}--\eqref{eq:dyn_beta} and:
\begin{align}  
\begin{split}
       \ddot{x}_t {\left(\bar{J}_x + m_t + m_r\right)} &=\, K_{x} u_x - \fricfunXbar  \\ & -\left(K_l u_l  -  \fricfunlbar \right)\sinalpha\,\sinbeta \\&+\tfrac{D_\alpha}{L}\dot{\alpha}\cosalpha\,\sinbeta + \tfrac{D_\beta}{L\sinalpha}\dot{\beta}\cosbeta\,,
\end{split}
\label{eq: x completa}
\end{align}
\begin{align}
\begin{split}
    \ddot{y}_t \left({\bar{J}_y + m_t}\right) &=\, K_{y} u_y -  \fricfunYbar   \\  &-\left(K_l u_l-  \fricfunlbar \right)\cosalpha - \tfrac{D_\alpha}{L}\dot{\alpha}\sinalpha\,,
\end{split}
\label{eq: y completa}
\end{align}
\begin{align}
\begin{split}
\ddot{L} \left(m_p+\bar{J}_l\right)&= \,K_l u_l -  \fricfunlbar - m_p \ddot{y}_t \cosalpha   + m_p L \dot{\alpha}^2  \\  &+ m_p L \dot{\beta}^2\sin^2{\!\alpha} - m_p \ddot{x}_t \sinalpha \,\sinbeta   
            \\ & +m_p g  \cosbeta \,\sinalpha  \,.
\end{split}
\label{eq: L completa}
\end{align}
where
\begin{equation}
    \bar{J}_{x} = \frac{r_g^2}{R_x^2}J_x\;\;,\;\;\bar{J}_{y} = \frac{r_g^2}{R_y^2}J_y\;\;,\;\;\bar{J}_{l} = \frac{r_g^2}{R_l^2}J_l\,,
    \label{eq: relaciones inercias radios}
\end{equation}
\begin{equation}
    K_{x} = \frac{r_g K_p}{R R_x}\;\;,\;\;K_{y} = \frac{r_g K_p}{R R_y}\;\;,\;\;K_{l} = \frac{r_g K_p}{R R_l}\,,
    \label{eq: relaciones Ks radios}
\end{equation}
and
\begin{equation}
    \fricfunXbar = \tfrac{K_p K_e r_g^2}{R R_x^2}\cdot\dot{x}_t +  \tfrac{r_g}{R_x}\cdot{\tau}_f\!\left(\tfrac{r_g}{R_x}\,\dot{x}_t\right) + \fricfunX \,, 
    \label{eq: def dx_barra}    
\end{equation}
\begin{equation}
    \fricfunYbar = \tfrac{K_p K_e r_g^2}{R R_y^2}\cdot\dot{y}_t +  \tfrac{r_g}{R_y}\cdot{\tau}_f\!\left(\dot{y}_t\tfrac{r_g}{R_y}\right) + \fricfunY \,,
    \label{eq: def dy_barra}
\end{equation}
\begin{flalign}
    &\;\;\fricfunlbar =  \tfrac{K_p K_e r_g^2}{R R_l^2}\cdot\dot{L} + \tfrac{r_g}{R_l}\cdot{\tau}_f\!\left(\dot{L}\tfrac{r_g}{R_l}\right) \,.
     \label{eq: def dl_barra} &&
\end{flalign}
are the effective functions that integrate the internal motor friction torques with the mechanical friction.

%----------------------------------------------------------------------------%
\section{Modeling and implementation of friction}
\label{sec:modelo friccion}
%----------------------------------------------------------------------------%
\subsection{Preliminary analysis}
\begin{figure}[t]
\centering
  \includegraphics[width=8.5cm]{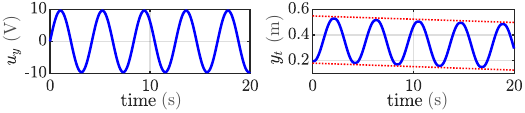}
  \caption{Experimental data from the real 3D-crane setup. The first plot shows the sinusoidal voltage input $u_y$ applied to the Y-axis motor, while the second plot displays the corresponding trolley position $y_t$ measured along the Y-axis. A clear drift is observed.}\label{fig:friccion_sentido}
\end{figure}

To characterize the system's friction, we experimentally investigated its dependence on position and direction of motion using our laboratory setup. First, a test conducted by applying a symmetric sinusoidal voltage to the Y-axis motor revealed a clear drift in the trolley's position response over time (as shown in Fig. \ref{fig:friccion_sentido}). This behavior indicates an asymmetry in the friction force, confirming its dependence on the direction of motion. 

Second, by measuring the voltage required to initiate movement at various points along the rail, we observed a clear dependence of friction on the position. These measures are described in detail in Section \ref{sec:resultados}. Based on these experimental findings, and in line with the friction modeling challenges highlighted in previous literature \cite{khatamianfar_new_2014,lobe_flatness-based_2018}, the remainder of this section introduces a comprehensive friction formulation. Specifically, to capture the observed real-world behavior, we propose the use of a viscous friction model that depends on the direction of motion, alongside a dry friction model that depends on both the position and the velocity of the system. 

\subsection{Friction model formulation}
Based on the experimental analysis, the friction models for the X- and rope-axes are described below. The Y-axis is omitted because it is equivalent to the X-axis.

For the trolley movement on the \mbox{X-axis} the mechanical friction force $f_x$ will be modeled as the sum of two terms,
\begin{equation}
    \fricfunX = D_x(\dot{x}_t)\,\dot{x}_t + C_x(x_t,\dot{x}_t)\;,
    \label{eq: def dx}
\end{equation}
where $D_x(\dot{x}_t)$ and $C_x(x_t,\dot{x}_t)$ are the direction-dependent viscous and the position- and direction-dependent dry friction functions, respectively.

On the other hand, the internal motor friction will be modeled as a position-independent torque:
\begin{equation}
    {\tau}_f(\dot{\theta}) = D_m\,\dot{\theta} + C_m\,\mathrm{sign}(\dot{\theta})\;,
    \label{eq: def dl}
\end{equation}
where $D_m$ and $C_m$ are the coefficients for viscous and dry friction respectively. 

\subsection{Friction components}
\subsubsection{Viscous friction}
The directional dependence of viscous friction on the X-axis is modeled using two distinct coefficients, ${D}_x^{\supermas}$ for positive velocity and ${D}_x^{\supermenos}$ for negative velocity:
 \begin{equation}
        {D}_x(\dot{x}_t) =
        \begin{cases}
          {D}_x^{\supermenos} & \text{if } \dot{x}_t < 0 \\
          0               & \text{if } \dot{x}_t = 0 \\
          {D}_x^{\supermas} & \text{if } \dot{x}_t > 0
        \end{cases} 
        \label{eq: dx sin barra}
        \end{equation}

\subsubsection{Dry friction} 
Dry friction is often ideally modeled as a force $F_s$ of constant magnitude that opposes motion, 
\begin{equation}
\label{eq:Fseca_con_sign}
    F_s = -C\,\mathrm{sign}(v) \;,
\end{equation}
where $v$ is the relative velocity, and $C$ is the dry friction coefficient.
The discontinuity of the $\mathrm{sign}$ function at $v=0$ in \eqref{eq:Fseca_con_sign} introduces a chattering problem on numerical simulation. A common solution is to approximate the $\mathrm{sign}$ function using the hyperbolic tangent ($\tanh$) \cite{wenzl_comparison_2018}, resulting in the following expression for dry friction:
\begin{equation}
    F_s = -C\tanh{(\gamma \cdot v)}   \;, 
\end{equation}
where $\gamma$ is a gain factor that adjusts the slope of the transition at zero-crossing. Increasing $\gamma$ reduces the approximation error, but at the cost of significantly increasing the simulation time. While this approximation may be valid for control purposes, on simulation it introduces errors that require high computation times to remain low.

Our model accounts for the dependence of dry friction on position and direction during the crane's movement along the X-axis, while maintaining the ideal discontinuity philosophy when changing direction, by means of a piecewise function:
 \begin{equation}
    {C}_x(x_t,\dot{x}_t) =
    \begin{cases}
      {C}_x^{\supermenos}(x_t)  & \text{if } \dot{x}_t < 0 \\
      0     & \text{if } \dot{x}_t = 0 \\
      {C}_x^{\supermas}(x_t)    & \text{if } \dot{x}_t > 0
    \end{cases}\;,
    \label{eq: c sin barra depende x y dx}
\end{equation}
where ${C}_x^{\supermenos}(x_t)$ and ${C}_x^{\supermas}(x_t)$ are the trolley position-dependent friction functions on the negative and positive directions, respectively.

\subsection{Effective friction}

The terms of the effective friction equations \eqref{eq: def dx_barra}--\eqref{eq: def dl_barra} can be regrouped as follows:
\begin{align}
   \fricfunXbar &=  \bar{D}_x(\dot{x}_t)\dot{x}_t + \bar{C}_x(x_t,\dot{x}_t) \;,
   \label{eq: dx_barra con pars} \\[1ex]
   \fricfunlbar &=  \bar{D}_l\dot{L} + \bar{C}_l\,\mathrm{sign}(\dot{L}) \;,
   \label{eq: dl_barra con pars}
\end{align}
where mechanical friction is neglected on the rope-axis model, including only the reflected motor effects.
From \eqref{eq: def dx}--\eqref{eq: dx sin barra}, the effective viscous friction components are obtained, as a piecewise constant function for $\bar{D}_x(\dot{x}_t)$ and as a constant for $\bar{D}_l$:
       \begin{equation}
        \bar{D}_x(\dot{x}_t) =
        \begin{cases}
          \bar{D}_x^{\supermenos} = \frac{K_pK_er_g^2}{RR_x^2}+\frac{r_g^2}{R_x^2}D_m+D_x^{\supermenos} & \text{if } \dot{x}_t < 0 \\
          0               & \text{if } \dot{x}_t = 0 \\
          \bar{D}_x^{\supermas} = \frac{K_pK_er_g^2}{RR_x^2}+\frac{r_g^2}{R_x^2}D_m+D_x^{\supermas} & \text{if } \dot{x}_t > 0
        \end{cases}\;, 
        \label{eq: dbarra_piecewise}    
        \end{equation}
        \begin{equation}
    \bar{D}_l = \tfrac{K_p K_e r_g^2}{R R_l^2}  + \tfrac{r_g^2}{R_l^2} D_{m} \;.
\end{equation}
Similarly, the effective dry friction components $\bar{C}_x(x_t,\dot{x}_t)$ and $\bar{C}_l$ are derived from \eqref{eq: def dx}--\eqref{eq: def dl} and \eqref{eq: c sin barra depende x y dx} as follows:
 \begin{equation}
    \bar{C}_x(x_t,\dot{x}_t) =
    \begin{cases}
      \bar{C}_x^{\supermenos}(x_t) = -\frac{r_g}{R_x}C_m+ {C}_x^{\supermenos}(x_t) & \text{if } \dot{x}_t < 0 \\
      0     & \text{if } \dot{x}_t = 0 \\
      \bar{C}_x^{\supermas}(x_t) =  +\frac{r_g}{R_x}C_m+ {C}_x^{\supermas}(x_t)  & \text{if } \dot{x}_t > 0
    \end{cases}
    \end{equation}
    \begin{equation}
    \bar{C}_l = \tfrac{r_g}{R_l} C_{m} \;.
\end{equation}

%----------------------------------------------------------------------------%
\subsection{Hybrid simulation model}
%----------------------------------------------------------------------------%

\begin{figure*}
\centering    
    \makebox[\textwidth][c]{\begin{tikzpicture}[
    >=latex, % Estilo de la punta de flecha 
    node distance=1.4cm and 0.4cm, % Distancia vertical y horizontal
    % Estilo de los bloques de estado
    stateblock/.style={
        rectangle, draw=black, thick, rounded corners=8pt,
        align=center, fill=cyan!15, % Color azul clarito
        text width=5.2cm, 
        minimum height=1.8cm,
        inner sep=1pt,
        font=\footnotesize\linespread{1.1}\selectfont 
    },
    % Estilo para los bloques del medio (más estrechos)
    stateblockmid/.style={
        stateblock,
        minimum height=1.2cm % Altura reducida
    },
    % Estilo del texto de las transiciones
    translabel/.style={
        align=center, font=\scriptsize, inner sep=2pt 
    }
]

% ==============================
% COLUMNA 1: X-AXIS
% ==============================
\node[stateblock] (X1) {
    \textbf{Mode q$_\mathbf{x}$=1:} Negative direction mov.\\
    $\ddot{x}_t(\bar{J}_x + m_t + m_r) = K_xu_x -\bar{D}_x^{\supermenos}\dot{x}_t + \bar{C}_x^{\supermenos}(x_t)
     - T_l\sinalpha\sinbeta +D_\alpha\dot{\alpha}\cosalpha\sinbeta/L + D_\beta\dot{\beta}\cosbeta/(L\sinalpha)$
};
\node[above=0cm of X1] {\textbf{X-AXIS}};

\node[stateblockmid, below=of X1] (X2) {
    \textbf{Mode q$_\mathbf{x}$=2:} No movement\\
    $\ddot{x}_t = 0$
};

\node[stateblock, below=of X2] (X3) {
    \textbf{Mode q$_\mathbf{x}$=3:} Positive direction mov.\\
    $\ddot{x}_t(\bar{J}_x + m_t + m_r) = K_xu_x -\bar{D}_x^{\supermas}\dot{x}_t - \bar{C}_x^{\supermas}(x_t)
     - T_l\sinalpha\sinbeta +D_\alpha\dot{\alpha}\cosalpha\sinbeta/L + D_\beta\dot{\beta}\cosbeta/(L\sinalpha)$
};

% Transiciones X-AXIS (Más pegadas al centro: xshift=0.6cm)
% \draw[->, thick] ([xshift=-0.6cm]X1.south) to[bend right=20] node[translabel, left] {$\dot{x}_t=0$\\or\\$x_t=x_{t_{\text{MIN}}}$} ([xshift=-0.6cm]X2.north);
% \draw[->, thick] ([xshift=0.6cm]X2.north) to[bend right=20] node[translabel, right] {$(K_xu_x-T_l\sin\alpha\sin\beta$\\$ +(D_\alpha\dot{\alpha}\cos{\alpha}\sin{\beta}$ \\ $+ D_\beta\dot{\beta}\sin{\alpha}\cos{\beta})/L)$ \\ $< -\bar{C}_x^{(-)}(x_t)$} ([xshift=0.6cm]X1.south);

% Transiciones X-AXIS
\draw[->, thick] ([xshift=-0.6cm]X1.south) to[bend right=20] 
    node[translabel, left, text=green!50!black, align=center] {
        $\dot{x}_t=0$ \\ or \\ $x_t=x_{t_{\text{MIN}}}$ \\ 
        \textcolor{red}{$\Rightarrow \dot{x}_t=0$}
    } ([xshift=-0.6cm]X2.north);

\draw[->, thick] ([xshift=0.6cm]X2.north) to[bend right=20] 
    node[translabel, right, text=green!50!black, align=center] {
        $(K_xu_x-T_l\sinalpha\sinbeta$ \\ 
        $+D_\alpha\dot{\alpha}\cosalpha\sinbeta/L$\\ 
        $+ D_\beta\dot{\beta}\cosbeta/(L\sinalpha)$ \\ 
        $< -\bar{C}_x^{\supermenos}(x_t)$ 
        % Si aquí añadieras un reset, sería: \\ \textcolor{red}{$\Rightarrow \dots$}
    } ([xshift=0.6cm]X1.south);

\draw[->, thick] ([xshift=-0.6cm]X2.south) to[bend right=20] 
    node[translabel, left, text=green!50!black, align=center] {$(K_xu_x-T_l\sinalpha\sinbeta$\\$ +D_\alpha\dot{\alpha}\cosalpha\sinbeta/L$ \\ $+ D_\beta\dot{\beta}\cosbeta/(L\sinalpha)$ \\ $> \bar{C}_x^{\supermas}(x_t)$} ([xshift=-0.6cm]X3.north);
\draw[->, thick] ([xshift=0.6cm]X3.north) to[bend right=20]
     node[translabel, right, text=green!50!black, align=center] {$\dot{x}_t=0$\\or\\$x_t=x_{t_{\text{MAX}}}$ \\ 
        \textcolor{red}{$\Rightarrow \dot{x}_t=0$}} ([xshift=0.6cm]X2.south);

% ==============================
% COLUMNA 2: Y-AXIS
% ==============================
\node[stateblock, right=1cm of X1] (Y1) {
    \textbf{Mode q$_\mathbf{y}$=1:} Negative direction mov.\\
    $\ddot{y}_t(\bar{J}_y + m_t) = K_yu_y -\bar{D}_y^{\supermenos}\dot{y}_t + \bar{C}_y^{\supermenos}(y_t)$\\
    $ - T_l\cosalpha - D_\alpha\dot{\alpha}\sinalpha/L$
};
\node[above=0cm of Y1] {\textbf{Y-AXIS}};

\node[stateblockmid, below=of Y1] (Y2) {
    \textbf{Mode q$_\mathbf{y}$=2:} No movement\\
    $\ddot{y}_t = 0$
};

\node[stateblock, below=of Y2] (Y3) {
    \textbf{Mode q$_\mathbf{y}$=3:} Positive direction mov.\\
    $\ddot{y}_t(\bar{J}_y + m_t) = K_yu_y-\bar{D}_y^{\supermas}\dot{y}_t - \bar{C}_y^{\supermas}(y_t)$ \\
    $ - T_l\cosalpha - D_\alpha\dot{\alpha}\sinalpha/L$
};

% Transiciones Y-AXIS
\draw[->, thick] ([xshift=-0.6cm]Y1.south) to[bend right=20] 
    node[translabel, left , text=green!50!black, align=center] {$\dot{y}_t=0$\\or\\$y_t=y_{t_{\text{MIN}}}$ \\ \textcolor{red}{$\Rightarrow \dot{y}_t=0$}} ([xshift=-0.6cm]Y2.north);
\draw[->, thick] ([xshift=0.6cm]Y2.north) to[bend right=20] 
node[translabel, right, text=green!50!black, align=center] {$(K_yu_y-T_l\cosalpha$\\$- D_\alpha\dot{\alpha}\sinalpha/L)$\\$ < -\bar{C}_y^{\supermenos}(y_t)$} ([xshift=0.6cm]Y1.south);

\draw[->, thick] ([xshift=-0.6cm]Y2.south) to[bend right=20] 
node[translabel, left, text=green!50!black, align=center] {$(K_yu_y-T_l\cosalpha$\\$- D_\alpha\dot{\alpha}\sinalpha/L)$\\$ > \bar{C}_y^{\supermas}(y_t)$} ([xshift=-0.6cm]Y3.north);
\draw[->, thick] ([xshift=0.6cm]Y3.north) to[bend right=20] 
node[translabel, right, text=green!50!black, align=center] {$\dot{y}_t=0$\\or\\$y_t=y_{t_{\text{MAX}}}$ \\ \textcolor{red}{$\Rightarrow \dot{y}_t=0$}} ([xshift=0.6cm]Y2.south);

% ==============================
% COLUMNA 3: ROPE-AXIS
% ==============================
\node[stateblock, right=1cm of Y1] (L1) {
    \textbf{Mode q$_\mathbf{l}$=1:} Negative direction mov.\\(ascending, \mbox{$\dot{L}<0$})\\
    $T_l = K_lu_l - \bar{D}_l\dot{L} + \bar{C}_l$\\
    $\ddot{L} \leftarrow \eqref{eq: L completa}$
};
\node[above=0cm of L1] {\textbf{ROPE-AXIS}};

\node[stateblockmid, below=of L1] (L2) {
    \textbf{Mode q$_\mathbf{l}$=2:} No movement\\
    $T_l  \leftarrow \eqref{eq:Tension_cuerda_parada}$\\
     $\ddot{L} = 0$
};

\node[stateblock, below=of L2] (L3) {
    \textbf{Mode q$_\mathbf{l}$=3:} Positive direction mov.\\(descending, \mbox{$\dot{L}>0$})\\
    $T_l = K_lu_l - \bar{D}_l\dot{L} - \bar{C}_l$\\
    $\ddot{L} \leftarrow \eqref{eq: L completa}$
};

% Transiciones ROPE-AXIS
\draw[->, thick] ([xshift=-0.6cm]L1.south) to[bend right=20] 
    node[translabel, left, text=green!50!black, align=center] {$\dot{L}=0$\\or\\$L=L_{\text{MIN}}$ \\ \textcolor{red}{$\Rightarrow \dot{L}=0$}} ([xshift=-0.6cm]L2.north);
\draw[->, thick] ([xshift=0.6cm]L2.north) to[bend right=20] 
    node[translabel, right, text=green!50!black, align=center] {$T_l + K_lu_l$ \\$< -\bar{C}_l$} ([xshift=0.6cm]L1.south);

\draw[->, thick] ([xshift=-0.6cm]L2.south) to[bend right=20]
     node[translabel, left , text=green!50!black, align=center] {$T_l + K_lu_l$ \\$> \bar{C}_l$} ([xshift=-0.6cm]L3.north);
\draw[->, thick] ([xshift=0.6cm]L3.north) to[bend right=20] 
    node[translabel, right, text=green!50!black, align=center] {$\dot{L}=0$\\or\\$L=L_{\text{MAX}}$ \\ \textcolor{red}{$\Rightarrow \dot{L}=0$}} ([xshift=0.6cm]L2.south);

\end{tikzpicture}}    
    \caption{Hybrid model of the 3 axis dynamics.}\label{hibrido_completo}    
\end{figure*}

To address the numerical issues caused by the dry friction simulation and to easily incorporate the dependence of friction on the direction of motion and position we propose the use of a hybrid simulation model \cite{goebel_hybrid_2009}. This model defines three distinct motion states for each axis of movement. These states are: negative direction motion, \mbox{no-motion}, and positive direction motion. Thus, the proposed hybrid dynamical model has 27 different dynamical modes, coded by the discrete variable $q$, defined as:
\begin{equation}
    q = [\;q_x \;\; q_y \;\; q_l\;]^\intercal \,,   
\end{equation}
where $q_x, q_y, q_l \in \{1,2,3\}^3$ describe the motion states of the X-, Y- and rope-axis respectively.

Dynamic modes are interconnected via \textit{guard} and \textit{reset} maps, forming the hybrid automaton shown in Fig.~\ref{hibrido_completo}. Transitions are defined by guard conditions (in \textcolor{green!50!black}{green}) and, when applicable, reset functions (in \textcolor{red}{red}) following the $\Rightarrow$ arrow. Motion initiates in each axis when the total force exceeds the corresponding position- and direction-dependent dry friction. Conversely, a transition to the rest state occurs when the velocity reaches zero or a physical position limit is encountered.

With this implementation, we account for the system behavior both when the rope length is changing and when $\dot{L} = 0$. In the first case, the rope tension $T_l$ is directly driven by the applied force minus friction. However, in the second case, $T_l$ requires accounting for the gravitational load, centrifugal effects, and the coupling between the cart's acceleration and the payload's oscillation, leading to the following expression:
\begin{equation}
\begin{split}
    \label{eq:Tension_cuerda_parada}
T_{l} = &\Big(-m_p \ddot{y}_t \cosalpha     + m_p L \dot{\alpha}^2  + m_p L \dot{\beta}^2\sin^2{\!\alpha}   \Big. \\
           &\Big. - m_p \ddot{x}_t \sinalpha\,\sinbeta  +  m_p g \cosbeta\,\sinalpha  \Big) \;.
\end{split}
\end{equation}
The physical limits on each axis are represented with the subscripts $\mathrm{MIN}$ and $\mathrm{MAX}$.

%%%%%%%%%%%%%%%%%%%%%%%%%%%%%%%%%%%%%%%%%%%%%%%%%%%%%%%%%%%%%%%%%%%%%%%%%%%%%%
\section{Simulation}
\label{sec:resultados_simulacion}
%%%%%%%%%%%%%%%%%%%%%%%%%%%%%%%%%%%%%%%%%%%%%%%%%%%%%%%%%%%%%%%%%%%%%%%%%%%%%%
This section presents two simplified simulation case studies to compare to demonstrate the superior speed and accuracy of our proposed hybrid model compared to the $\tanh$ approximation. To isolate the effects of the dry friction model which is the key feature of the hybrid approach, these simulations are performed without viscous friction ($\bar{D}_x = \bar{D}_y = \bar{D}_l = 0$) and motor inertia ($J=0$). The remaining physical parameters not explicitly detailed are not critical for these case studies, as the objective is to demonstrate the qualitative behavior. For simplicity, dry friction is assumed constant across all positions and directions in each axis. Simulations were performed in MATLAB using an Intel Ultra 7 265K (3.90 GHz) with 32 GB of RAM.

\begin{figure}[t]
\centering
  \includegraphics[width=8cm]{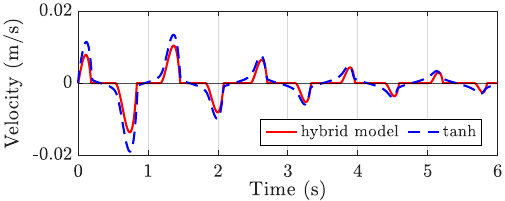}
  \caption{Case 1 comparison. The plot shows the trolley velocity with the hybrid model and the continuous model based on the $\tanh$ function with $\gamma = 1000$ respectively.}\label{caso1}

\vspace{\floatsep}

\centering
  \includegraphics[width=8cm]{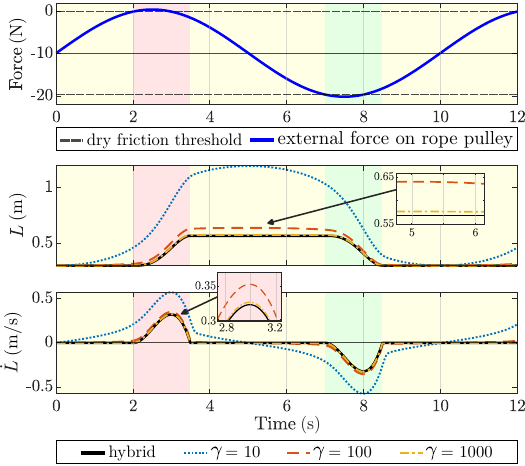}
  \caption{Case 2 comparison. The top plot illustrates the forces acting on the rope axis with the hybrid model. The middle plot displays the rope length and the bottom plot compares the rope length velocity obtained from the hybrid model and the $\tanh$ model with varying $\gamma$ values. Colored backgrounds denote negative motion (red), rest (yellow), and positive motion (green) on the hybrid model.}\label{caso3}
\end{figure}

\begin{itemize}
    \item \textbf{Case 1.} A simulation is performed with a payload mass of 0.457 kg and a dry friction force of $\bar{C}_x = \bar{C}_y =1 $N. The system is released from an initial position $\alpha(0) = \frac{\pi}{2} + 0.6$ rad, $\beta(0) = 0$ rad, with no external forces ($u_x=u_y=u_l=0$) and constant rope length. The force transmitted from the payload sway to the trolley causes it to move once the dry friction threshold is overcome.
    Fig. \ref{caso1} shows that the $\tanh$ model never allows the trolley to come to a complete stop, whereas our hybrid model correctly captures that effect, where the trolley remains stationary when the net force is below the dry friction limit. 
    \item \textbf{Case 2.} This case considers motion only along the rope axis, with a payload mass of 1 kg and a dry friction coefficient of $\bar{C}_l = 9.81$ N. A sinusoidal force $F_l$ is applied to the rope axis. The results are shown in Fig. \ref{caso3}.  
    It can be observed that the rope length varies only when the applied force exceeds the corresponding dry friction threshold. The bottom plot shows the velocity for various $\tanh$ models compared to the hybrid model reference.
\end{itemize}
Table \ref{tab:comparativacasos} further highlights the computational advantage of the hybrid approach in both cases. It is shown that while the $\tanh$ model may be faster for large approximation errors, the hybrid model achieves significantly lower simulation times when high precision is required.

\renewcommand{\arraystretch}{1.3}
\begin{table}[h]
    \centering
    \setlength{\tabcolsep}{5pt}
    \caption{Comparison of simulation time between the $\tanh$ dry friction modeling and our hybrid model.}
    \begin{tabular}{c|cc|cc}
        \hline
        Model & \multicolumn{2}{c|}{\textbf{CASE 1}} & \multicolumn{2}{c}{\textbf{CASE 2}}\\
        \hline
         & $t_{sim}$ (s)& $RMSE$ (m/s)& $t_{sim}$ (s)& $RMSE$ (m/s)\\
         \hline
         hybrid & 0.216 & 0 & 0.0658 & 0 \\
         $\tanh(v)$ & 0.165 & 0.0864 & 0.0933 & 1.19 \\
         $\tanh(10 v)$ & 0.276 & 0.0327 & 0.229 & 0.148 \\
         $\tanh(100 v)$ & 1.235 & 0.00821 & 0.572 & 0.0174 \\
         $\tanh(1000 v)$ & 11.824 & 0.00168 & 1.917 & 0.00201 \\
         $\tanh(10000 v)$ & 117.726 & 0.00102 & 18.219 & 0.000228 \\
         \hline        
    \end{tabular}
    \label{tab:comparativacasos}
\end{table}
\renewcommand{\arraystretch}{1}

%%%%%%%%%%%%%%%%%%%%%%%%%%%%%%%%%%%%%%%%%%%%%%%%%%%%%%%%%%%%%%%%%%%%%%%%%%%%%%
\section{Parameter estimation}
\label{sec:identif}
%%%%%%%%%%%%%%%%%%%%%%%%%%%%%%%%%%%%%%%%%%%%%%%%%%%%%%%%%%%%%%%%%%%%%%%%%%%%%%

Based on the hybrid model equations, 14 parameters, $\bar{J}_l$, $K_l$, $\bar{C}_l$, $\bar{D_l}$, $\bar{J}_x$, $K_x$, $\bar{D}_x^{\supermas}$, $\bar{D}_x^{\supermenos}$, $\bar{J}_y$, $K_y$, $\bar{D}_y^{\supermas}$, $\bar{D}_y^{\supermenos}$, $D_\alpha$, $D_\beta$, and 4 functions, $\bar{C}_x^{\supermas}(x_t)$, $\bar{C}_x^{\supermenos}(x_t)$,  $\bar{C}_y^{\supermas}(y_t)$, $\bar{C}_y^{\supermenos}(y_t)$, need to be estimated, while the remaining constants (masses, pulley radii, and $r_g$) can be directly measured.
Our estimation strategy, inspired by the method in \cite{khatamianfar_new_2014}, proposes a multi-step method that separates the estimation of dry friction functions using Bayesian Linear Regression from the LS estimation of the parameters.
In this type of systems, it is common for the physical characteristics of the motors to be known. In our setup, however, this is not the case. We do, however, consider the three DC motors to be identical, allowing for a no-load estimation using data from the rope-axis motor when it is not carrying a load.

The complete estimation process is conducted by first estimating the parameters of the motor and the rope axis through tests on that axis, then estimating the parameters of the other two axes separately and finally estimating $D_\alpha$ and $D_\beta$.

%----------------------------------------------------------------------------%
\subsection{Rope-axis parameters estimation}
%----------------------------------------------------------------------------%

For the parameter estimation of the rope-axis, we use data from tests where only the rope is in motion and the payload is not oscillating ($\alpha = \pi/2$, $\beta = 0$). Under these conditions, \eqref{eq: L completa} and \eqref{eq: dl_barra con pars} simplify into:
\begin{equation}
\label{eq: cuerda identificacion ps}
    u_l = P_1\ddot{L}+P_2\dot{L}+P_3\mathrm{sign}(\dot{L})+P_4m_p(\ddot{L}-g)\;,
\end{equation}
where, to facilitate the estimation process, $P_1$, $P_2$, $P_3$ and $P_4$ are auxiliary estimation parameters defined as
\begin{equation}
    \label{eq: relacionesP_params1}
    P_1 = \frac{\bar{J_l}}{K_l} \;\;,\;\; P_2 = \frac{\bar{D_l}}{K_l} \;\;,\;\; P_3 = \frac{\bar{C_l}}{K_l} \;\;,\;\; P_4 = \frac{{1}}{K_l}\;.
\end{equation}

%-.-.-.-.-.-.-.-.-.-.-.-.-.-.-.-.-.-.-.-.-.-.-.-.-.-.-.-.-.-.-.-.-.-.-.-.-.%
\subsubsection{No-load estimation (quasistatic)}
%-.-.-.-.-.-.-.-.-.-.-.-.-.-.-.-.-.-.-.-.-.-.-.-.-.-.-.-.-.-.-.-.-.-.-.-.-.%
\label{subsub: cuerda quasistatic}
In a first stage, the motor displacement without payload ($m_p = 0$) is measured.
Starting from rest ($\dot{L} = \ddot{L} = 0$), a slow voltage ramp is applied to find the exact moment the motion begins and to estimate $P_3$ via the quasistatic relation
\begin{equation}
    u_l = P_3\;\mathrm{sign}(\dot{L})\;.
\end{equation}

%-.-.-.-.-.-.-.-.-.-.-.-.-.-.-.-.-.-.-.-.-.-.-.-.-.-.-.-.-.-.-.-.-.-.-.-.-.%
\subsubsection{No-load estimation (movement)}
%-.-.-.-.-.-.-.-.-.-.-.-.-.-.-.-.-.-.-.-.-.-.-.-.-.-.-.-.-.-.-.-.-.-.-.-.-.%
\label{subsub: cuerda mov sin carga}
$P_1$ and $P_2$ can be estimated from a test with movement in the rope-axis, using a standard LS method \cite{ljung_system_1999} based on the following linear regression model:
\begin{equation}
    u_l = P_1\;\ddot{L}+P_2\;\dot{L}+P_3\;\mathrm{sign}(\dot{L})\;.
\end{equation}

%-.-.-.-.-.-.-.-.-.-.-.-.-.-.-.-.-.-.-.-.-.-.-.-.-.-.-.-.-.-.-.-.-.-.-.-.-.%
\subsubsection{Loaded movement estimation}
%-.-.-.-.-.-.-.-.-.-.-.-.-.-.-.-.-.-.-.-.-.-.-.-.-.-.-.-.-.-.-.-.-.-.-.-.-.%
\label{subsub: cuerda mov con carga}
For this step, experiments are conducted with a non-zero payload. Using the previously estimated motor parameters, and \eqref{eq: cuerda identificacion ps}, $P_4$ is estimated via LS.
A separate estimation for $P_4$ is required because its associated regressor is highly correlated with that of $P_1$ (as both depend on acceleration), which would otherwise lead to a poorly conditioned estimation.

Having obtained $P_1$, $P_2$, $P_3$ and $P_4$, $\bar{J_l}$, $K_l$, $\bar{D_l}$ and $\bar{C_l}$ are calculated using \eqref{eq: relacionesP_params1}.

%----------------------------------------------------------------------------%
\subsection{X and Y axes motion estimations}
%----------------------------------------------------------------------------%

The dry friction functions and motion parameters are estimated from unloaded motion tests ($m_p=0$) for each axis. By simplifying \eqref{eq: x completa} and applying \eqref{eq: dx_barra con pars}--\eqref{eq: dbarra_piecewise}, the following relation is obtained:
\begin{equation}
\label{eq: ux motion}
    u_x\;K_x = (\bar{J_x} + m_t+m_r)\ddot{x}_t+\bar{D}_x(\dot{x}_t)\dot{x}_t + \bar{C}_x(x_t, \dot{x}_t)\;,
\end{equation}
where $K_x$ and $\bar{J_x}$ can be derived from the previously estimated parameters $K_l$ and $\bar{J_l}$ using the radii relations given in \eqref{eq: relaciones inercias radios} and \eqref{eq: relaciones Ks radios}.

%----------------------------------------------------------------------------%
\subsubsection{Dry friction functions estimation}
%----------------------------------------------------------------------------%

\label{subsub: xy coulomb}
The dry friction terms for the X and Y axes are estimated independently. To model the position-dependent dry friction, various analytical models of varying complexity could be employed. However, based on empirical observations and to avoid introducing unnecessary computational complexity during simulation, we fit 4th-order polynomials to the experimental data:
    \begin{equation}
    \begin{split}
       \bar{C}_x^{\supermasmenos}(x_t) = c_{x0}^{\supermasmenos}+c_{x1}^{\supermasmenos}x_t+c_{x2}^{\supermasmenos}x_t^2+c_{x3}^{\supermasmenos}x_t^3+c_{x4}^{\supermasmenos}x_t^4 \;.
       \label{eq: superficie_friccion_menos}
    \end{split}        
    \end{equation}
   
Since exhaustive measurement across the entire workspace is impractical, an active learning approach based on Bayesian Linear Regression \cite{bishop2006pattern} is employed to identify the friction polynomials. Unlike deterministic curve-fitting, this probabilistic framework estimates the posterior distribution of the polynomial coefficients, yielding both the expected friction and its predictive variance for any unmeasured location across the axis.

To minimize the number of physical experiments, data acquisition is performed sequentially. The process starts with a sparse initial dataset, which is then iteratively expanded following these steps:
\begin{enumerate}
    \item \textbf{Posterior Update:} The current dataset is used to update the mean and covariance of the polynomial coefficients.
    \item \textbf{Optimal Point Selection:} An acquisition score is evaluated over a fine grid of candidate locations to determine the most informative next measurement. This score is constructed by multiplying the model's predictive standard deviation by the squared distance to the nearest previously measured point. This naturally balances the exploration of highly uncertain regions with a spatial penalty that ensures a well-distributed coverage of the workspace.
    \item \textbf{Measurement:} The trolley is moved to the location that maximizes the acquisition score. A voltage ramp is applied to identify the static friction components $\bar{C}_{x}^{\supermasmenos}$ and $\bar{C}_{y}^{\supermasmenos}$, which are then added to the dataset.
\end{enumerate}
This cycle repeats until the maximum predictive variance falls below a desired threshold, a maximum number of evaluation points is reached, or a predefined time limit expires.

%-.-.-.-.-.-.-.-.-.-.-.-.-.-.-.-.-.-.-.-.-.-.-.-.-.-.-.-.-.-.-.-.-.-.-.-.-.%
\subsubsection{Viscous friction coefficients estimation}
%-.-.-.-.-.-.-.-.-.-.-.-.-.-.-.-.-.-.-.-.-.-.-.-.-.-.-.-.-.-.-.-.-.-.-.-.-.%
\label{subsub: xy param}
The remaining dynamic parameters correspond to the viscous friction of the system.
Using the previously determined dry friction functions, a least squares regression is applied to \eqref{eq: ux motion}. This estimation yields the direction-dependent viscous friction coefficients $\bar{D}_{x}^{\supermasmenos}$.

The process for the Y-axis is completely equivalent.
%----------------------------------------------------------------------------%
\subsection{Angles viscous friction}
%----------------------------------------------------------------------------%
\label{sub: ang viscosa}

The last parameters to be estimated are the angles damping coefficients, $D_\alpha$ and $D_\beta$, associated with the payload swing.
To estimate these coefficients, we use data recorded while the payload is freely oscillating. To achieve that, the trolley and the rope are moved along a trajectory and then stopped. Thus,  $\ddot{x}_t = \dot{x}_t = \ddot{y}_t = \dot{y}_t = \ddot{L} = \dot{L} = 0$ during this experiment. 
Under these conditions, \eqref{eq:dyn_alpha} and \eqref{eq:dyn_beta} are simplified into:
\begin{align}
   L\;\ddot{\alpha} &= g \cosalpha  \cosbeta  + \cosalpha \sinalpha L {\dot{\beta}}^2 - \frac{D_{\alpha}}{m_pL}\,\dot{\alpha}\,, \\
    L\;\ddot{\beta} &=  \frac{-g \sinbeta  -2 \cosalpha  L \dot{\alpha}\dot{\beta}}{\sinalpha }  - \frac{D_\beta}{m_pL}\,\dot{\beta}\,,
\end{align}
and the damping coefficients can be directly estimated via LS.

\begin{figure}[t]
\centering
  \includegraphics[width=\columnwidth]{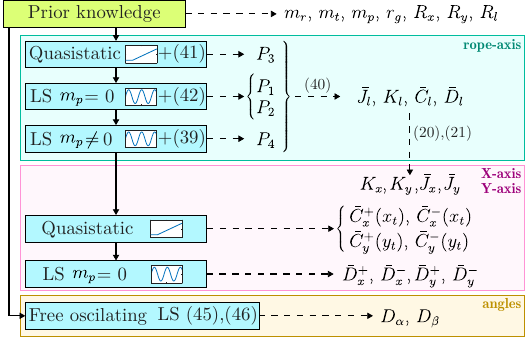} 
  \caption{Flowchart of the parameter estimation process. First, rope-axis parameters ($\bar{J}_l$, $K_l$, $\bar{C}_l$, $\bar{D}_l$) are estimated. These are then used to calculate $K_x$, $K_y$, $\bar{J}_x$ and $\bar{J}_y$, and the rest of the estimation for the X- and Y-axis is performed. 
  Independently, $D_\alpha$, and $D_\beta$ are obtained.}\label{fig:esquema_identif}
\end{figure}

A flowchart of the complete process is shown in Fig. \ref{fig:esquema_identif}.

%%%%%%%%%%%%%%%%%%%%%%%%%%%%%%%%%%%%%%%%%%%%%%%%%%%%%%%%%%%%%%%%%%%%%%%%%%%%%%
\section{Experimental results}
\label{sec:resultados}
%%%%%%%%%%%%%%%%%%%%%%%%%%%%%%%%%%%%%%%%%%%%%%%%%%%%%%%%%%%%%%%%%%%%%%%%%%%%%%
In this section, the experimental results of the complete estimation process and the comparison between the hybrid simulation model and the real data are discussed.
%----------------------------------------------------------------------------%
\subsection{Experimental setup description}
%----------------------------------------------------------------------------%

\begin{figure}[t]
\centering
  \includegraphics[width=8.5cm]{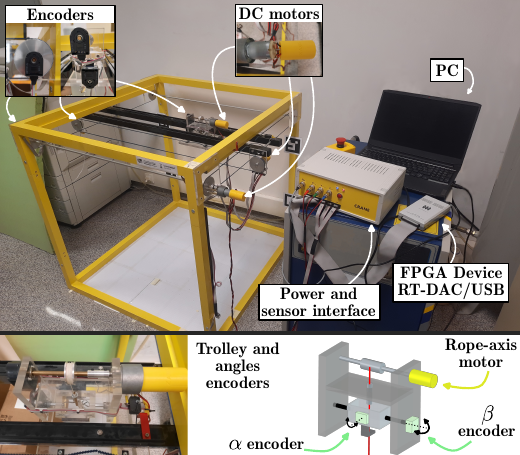}
  \caption{Experimental overhead crane setup. }\label{fig: foto_puente1}
\end{figure}

The 3D overhead crane used in this study is a laboratory setup composed by the INTECO 3D-Crane (see Fig. \ref{fig: foto_puente1}) and a personal computer (AMD Ryzen 5 3.301GHz CPU and 16GB of RAM). The crane is equipped with three 12V DC motors and five encoders, which provide measurements of the system outputs with a resolution of 4096 pulses per rotation and an accuracy of 0.0015 rad. Communication between the crane and the computer is established via the RT-DAC/USB2 input/output board from the same manufacturer. The programming of the board is performed using Matlab/Simulink and the dedicated toolbox of the manufacturer. The voltage applied to the DC motors is controlled through a $\pm12$ V PWM signal. 
The known physical parameters include the drive radii for the X, Y, and rope axes ($R_x = R_y = 40\cdot 10^{-3}$ m, $R_l = 15\cdot 10^{-3}$ m). The operational workspace is constrained by the mechanical limits, with positions bounded by $x_t \in [0, 0.505]$ m, $y_t \in [0, 0.625]$ m and $L \in [0.13, 0.57]$ m.

%----------------------------------------------------------------------------%
\subsection{Data processing}
%----------------------------------------------------------------------------%

The estimation of velocity and acceleration from position signals is affected by significant quantization noise from the crane's encoders. This noise level invalidates results from direct numerical differentiation. To address this, we employ a multi-step offline filtering and differentiation process.

First, to mitigate the effects of quantization noise, the raw position signal with a sample time of $T_s = 0.002$ seconds is processed with a zero-phase filter. We use a fourth-order Butterworth low-pass filter with 3 Hz cutoff frequency, applied forwards and then backwards to ensure there is no time delay introduced in the filtered signal.
From the filtered signal at instant $k$, ${x_{f}}_k$, the estimated velocity, $\hat{\dot{x}}$, and acceleration $\hat{\ddot{x}}$ are calculated using central differences:
\begin{equation}
    \hat{\dot{x}} = \frac{{x_{f}}_{k+1} - {x_{f}}_{k-1}}{2 \; T_s} \;,\;  \hat{\ddot{x}} = \frac{{x_{f}}_{k+1} - 2 \;{x_{f}}_k+{x_{f}}_{k-1}}{{T_s}^2} \;.
\end{equation}

\begin{figure}[t]
\centering
  \includegraphics[width=8.5cm]{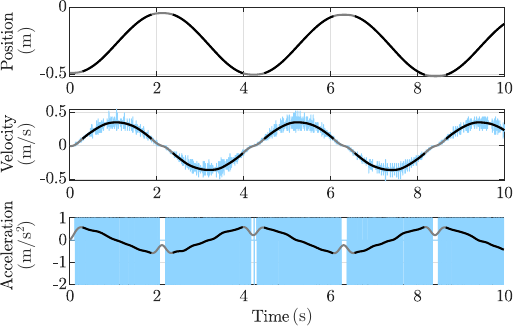}
  \caption{Example of signal processing for estimation. Gray lines: complete filtered signals. Black lines: data segments used for LS estimation after removing regions around zero-velocity crossings. Blue lines: noisy signals resulting from direct differentiation of raw data.} 
  \label{fig:friccion_trozos}
\end{figure}

A compromise between signal smoothness and dynamic accuracy was required when filtering, particularly near zero-velocity crossings where numerical differentiation fails due to dry friction. Consequently, filter parameters were optimized for smoothness, and poorly estimated regions were excluded from the identification process to ensure data quality. Fig. \ref{fig:friccion_trozos} shows an example of the calculation of velocity and acceleration from the recorded data. It shows how, without prior filtering, directly deriving the signal (blue lines) does not provide good data for parameter estimation.

%----------------------------------------------------------------------------%
\subsection{Results}
%----------------------------------------------------------------------------%

\begin{figure}[t]
\centering
  \includegraphics[width=8.5cm]{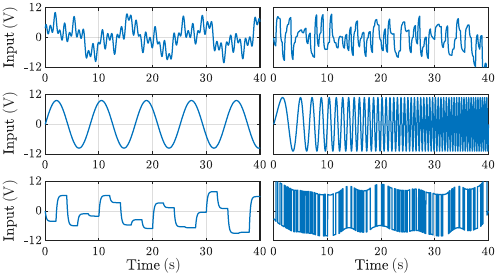}
  \caption{Input signals used for estimation purposes.}\label{fig: sigs_entrada}
\end{figure}

All data is sampled at $T_s = 0.002$ s. Each axis is excited with various signals (Fig. \ref{fig: sigs_entrada}), such as multi-steps and chirps (varying from 0.1 to 2.5 Hz). Amplitudes and frequencies are chosen to maximize trajectory richness within safe operational bounds, preventing data saturation at physical limits.

\begin{figure}[t]
\centering
  \includegraphics[width=8.5cm]{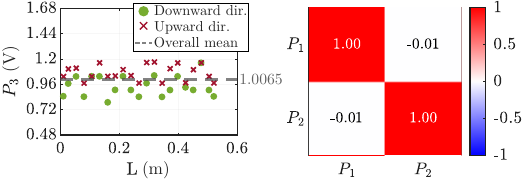}
  \caption{Absolute measurements of the voltage required to initiate motion ($P_3$), showing values for both directions and their overall mean (left). Correlation matrix from the LS estimation of parameters $P_1$ and $P_2$ (right).}\label{fig: iden_sin_carga}
\end{figure}

First, the motor's dry friction parameter $P_3$ is estimated via the quasistatic test (Section \ref{subsub: cuerda quasistatic}). As shown in Fig.~\ref{fig: iden_sin_carga} (left), $P_3$ exhibits negligible dependence on the motion direction; thus, a single mean value is adopted. Subsequently, $P_1$ and $P_2$ are estimated (Section \ref{subsub: cuerda mov sin carga}).
The LS correlation matrix \cite{keesman2011system} in Fig.~\ref{fig: iden_sin_carga} (right) reveals near zero off-diagonal values, confirming uncorrelated estimates.

Next, under load conditions (Section \ref{subsub: cuerda mov con carga}), data from tests with four different masses (0.064, 0.325, 0.457, and 0.550 kg) were concatenated to obtain a robust estimate of $P_4$.

\begin{figure}[t]
\centering
  \includegraphics[width=8.5cm]{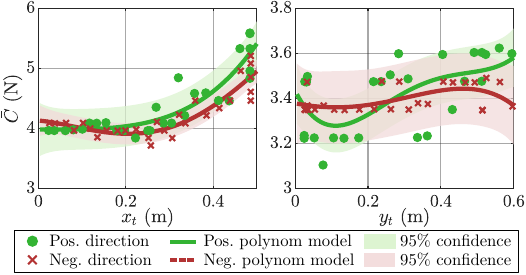}
  \caption{Absolute measurements of the force required to initiate motion, the polynomial models and the 95\% confidence bands (±1.96 standard deviations), showing the uncertainty of the estimation on X and Y axis.}\label{fig: gpr_juntas}
\end{figure}

Subsequently, the position- and direction-dependent dry friction for the X- and Y-axes is estimated. To properly configure the estimation algorithm, the measurement noise variances are determined via repeated measurements at multiple locations along each axis. The resulting variance values used for the positive and negative directions of the X-axis are 0.026 and 0.0073 N$^2$, respectively. Similarly, for the Y-axis, the obtained variances are 0.0030 and 0.0056 N$^2$.
Results from the methodology proposed in Section \ref{subsub: xy coulomb} are shown in Fig.~\ref{fig: gpr_juntas}.

To complete the estimation, unloaded dry friction parameters are estimated (Section \ref{subsub: xy param}) using the input signals from Fig.~\ref{fig: sigs_entrada}. 

Finally, swing angle damping coefficients ($D_\alpha$, $D_\beta$) are estimated from three free-oscillation tests with a 0.457 kg payload (Section \ref{sub: ang viscosa}). 
Table \ref{tab:resultados_estimacion} shows the results of the complete estimation process.

\renewcommand{\arraystretch}{1.2} 
\begin{table}[h!]
    \centering
    \caption{Estimated System Parameters and Dry Friction Coefficients.}
    \label{tab:resultados_estimacion}    
    
    \begin{tabular}{c|cc|c|cc}
        \hline        
        \textbf{Axis} & \textbf{Inertia} & \textbf{Gain} & \textbf{Dry Fric.} & \multicolumn{2}{c}{\textbf{Viscous Fric.}} \\        
         & $\bar{J}$ & $K$ & $\bar{C}^{\supermasmenos}$ & $\bar{D}^{\supermas}$ & $\bar{D}^{\supermenos}$\\
        \hline       
        $l$ & 18.19 & 8.53 & 8.583 & \multicolumn{2}{c}{701.81} \\
        $x$ & 2.56 & 3.20 & (See below) & 105.87 & 101.44\\
        $y$ & 2.56 & 3.20 & (See below) & 98.50 & 97.13\\
        $\alpha$ & --- & ---  & --- & \multicolumn{2}{c}{2.573e-04} \\
        $\beta$ & --- & ---  & --- & \multicolumn{2}{c}{0.0059} \\
        \hline
    \end{tabular}

    \vspace{0.3cm}

   \centering
    \setlength{\tabcolsep}{2.5pt}  
    \begin{tabular}{ccccccccccc}
        \hline
        & $c_{0}^{\supermas}$ & $c_{1}^{\supermas}$ & $c_{2}^{\supermas}$ & $c_{3}^{\supermas}$ & $c_{4}^{\supermas}$ & $c_{0}^{\supermenos}$ & $c_{1}^{\supermenos}$ & $c_{2}^{\supermenos}$ & $c_{3}^{\supermenos}$ & $c_{4}^{\supermenos}$ \\
        \hline
        $x$ & 1.23 & 0.29 & -1.95 & 3.98 & 5.02 & 1.29 & -0.089 & -4.24 & 17.82 & -13.75 \\
        $y$ & 1.08 & -0.95 & 6.08 & -12.41 & 8.56 & 1.06 & -0.09 & 0.29 & 0.79 & -1.74 \\
        \hline
    \end{tabular}
\end{table}
\renewcommand{\arraystretch}{1}

%----------------------------------------------------------------------------%
\subsection{Model validation.}
%----------------------------------------------------------------------------%

To validate the effectiveness and accuracy of the proposed hybrid model and the parameter estimation method, we present an experiment in which the results obtained from the laboratory setup are compared against those obtained with the simulator using identical input signals. In this experiment, a 0.064 kg payload is used.

The input signals, which differ from those used during the parameter estimation phase, consist of superimposed sine waves of varying frequencies and amplitudes. Specifically, the frequencies applied are 0.135, 0.404, 0.673, and 0.942 Hz for the X-axis; 0.105, 0.314, 0.523, and 0.942 Hz for the Y-axis; and 0.135, 0.406, 0.219, and 0.744 Hz for the rope-axis.

\begin{figure}[t]
\centering
  \includegraphics[width=8.5cm]{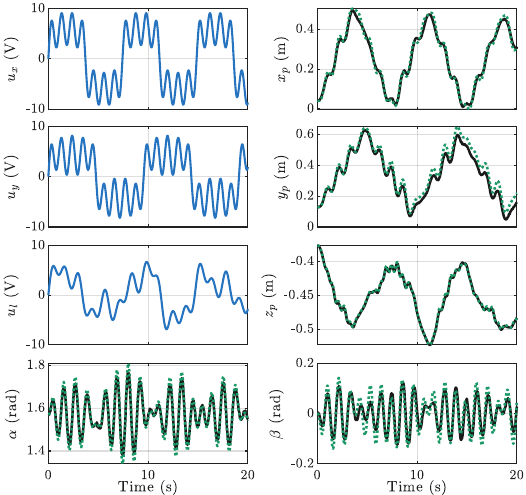}
  \caption{Comparison of measured (black) and simulated (dotted green) angles and payload position using the blue input signals.}\label{fig: validacion_multisine}
\end{figure}

Fig. \ref{fig: validacion_multisine} presents the input signals alongside a comparison of the payload position across the three axes and the swing angles. The results demonstrate high fidelity, with the outputs of the simulator closely matching the experimental data, thereby confirming the accuracy of the model in replicating the behavior of the system under these conditions. 
To evaluate the tracking performance, the Normalized Mean Absolute Error (NMAE) is calculated using the total physical range of each axis as the normalization factor, yielding values of 2.59\%, 4.95\%, and 0.13\% for the $x_c$, $y_c$, and $z_c$ coordinates, respectively.

Regarding the Y-axis payload position, although the simulated signal correctly reproduces the overall trajectory shape, a cumulative discrepancy can be observed. This deviation stems from minor inaccuracies in the highly sensitive dry friction modeling, which causes the trolley to start or stop slightly earlier or later than the real system.

%%%%%%%%%%%%%%%%%%%%%%%%%%%%%%%%%%%%%%%%%%%%%%%%%%%%%%%%%%%%%%%%%%%%%%%%%%%%%%
\section{Conclusions}
\label{sec:conclusiones}
%%%%%%%%%%%%%%%%%%%%%%%%%%%%%%%%%%%%%%%%%%%%%%%%%%%%%%%%%%%%%%%%%%%%%%%%%%%%%%

This article proposes a novel hybrid simulation model and a comprehensive parameter estimation methodology to accurately capture complex friction dynamics in overhead cranes simulations. 
By modeling friction dependencies on movement direction and trolley position through distinct states and transition conditions, our approach overcomes the precision and computational limitations of conventional models.
For reproducibility, the model is available on GitHub as an open-source benchmark.
Our step-by-step parameter estimation combines Least Squares (LS) with probabilistic dry friction fitting. Because this method inherently quantifies estimation uncertainty, it provides a solid foundation for introducing realistic stochasticity into simulators and for developing robust control strategies.

This work highlights that accounting for dry friction in the model is crucial for accurately describing its dynamics and, subsequently, for controlling the system’s movements. The model proposed here can be used, for example, to design feedforward controllers that enable the system to follow aggressive trajectories.

Despite these advantages, the proposed approach has inherent limitations. Representing dry friction as a position-dependent polynomial assumes smooth variation across the workspace. Consequently, the model cannot capture discrete, localized irregularities, such as structural cracks. To address this, future research will explore integrating neural networks to model these highly nonlinear friction phenomena.

%%%%%%%%%%%%%%%%%%

\section*{Acknowledgments}

The authors would like to thank Ram\'on Piedrafita for his time and for allowing us to use the overhead crane setup. The authors used \mbox{GEMINI} for language refinement, thoroughly reviewed the edited text, and assume full responsibility for the final manuscript.

\bibliographystyle{IEEEtran}


\end{document}